\documentclass[a4paper,12pt]{article}%
\usepackage{amsfonts}
\usepackage{amsmath}
\usepackage{amssymb}
\usepackage{graphicx}
\usepackage[all]{xy}
\usepackage{hyperref}
\usepackage{color}

\newcommand\on[1]{\operatorname{#1}}
\newcommand\mc[1]{\mathcal{#1}}
\newcommand\ps[1]{\underline{#1}}

\newcommand\De{\Delta}

\newcommand{\op}{\on{op}}              

\newcommand{\hP}{\hat{P}}

\newcommand\Ain[1]{A\,\varepsilon\,#1}

\newcommand{\Sig}{\ps{\Sigma}}            

\newcommand\Set{\mathbf{Set}}                    
\newcommand\cN{\mc{N}}

\newcommand\VN{\mc{V}(\cN)}
\newcommand\SetC[1]{\Set^{#1^{\op}}}
\newcommand\SetVNop{\SetC{\VN}}

\newcommand\bbR{\mathbb{R}}

\newcommand\wpsi{\ps{\mathfrak{w}^\psi}}

\newcommand\ra{\rightarrow}

\newcommand\lra{\longrightarrow}
\newcommand\lmt{\longmapsto}

\newcommand\Ga{\Gamma}

\newcommand\deo{\delta^o}

\newcommand\Rlr{\ps{\bbR^{\leftrightarrow}}}

\newcommand\cS{\mc S}

\begin{document}

\title{\textbf{Some Remarks on the Logic\\of Quantum Gravity}}
\author{Andreas D\"oring\\
\small{Clarendon Laboratory, Department of Physics, University of Oxford}\\
\normalsize{doering@atm.ox.ac.uk}}%
\date{\small{13. June 2013}}

\maketitle

\vspace{-0.8cm}

\begin{abstract}
We discuss some conceptual issues that any approach to quantum gravity has to confront. In particular, it is argued that one has to find a theory that can be interpreted in a realist manner, because theories with an instrumentalist interpretation are problematic for several well-known reasons. Since the Hilbert space formalism almost inevitably forces an instrumentalist interpretation on us, we suggest that a theory of quantum gravity should not be based on the Hilbert space formalism. We briefly sketch the topos approach, which makes use of the internal logic of a topos associated with a quantum system and comes with a natural (neo-)realist interpretation. Finally, we make some remarks on the relation between system logic and metalogic.
\end{abstract}


\noindent\textit{Should storms, as well may happen\\
\indent Drive you to anchor a week\\
In some old harbour-city\\
\indent Of Ionia, then speak\\
With her witty scholars, men\\
Who have proved there cannot be\\
\indent Such a place as Atlantis:\\
\indent Learn their logic, but notice\\
How its subtlety betrays\\
\indent Their enormous simple grief;\\
Thus they shall teach you the ways\\
\indent To doubt that you may believe.}

W.H. Auden, from \textit{Atlantis} (1941)

\vspace{0.8cm}

\noindent\textit{``Die Grenzen meiner Sprache bedeuten die Grenzen meiner Welt.''}

Ludwig Wittgenstein, \textit{Tractatus logico-philosophicus}, Satz 5.6

\section{Introduction}
Doing research in quantum gravity is a profoundly strange endeavour: neither the boundaries of the subject, nor the methods of inquiry, nor the goals of the search, nor the criteria of success are commonly agreed upon. This holds in particular when one considers not a specific approach with its often formidable technical apparatus and mathematical difficulties, but conceptual questions that are common to the various approaches.

There are no observable phenomena that unambiguously belong to the realm of quantum gravity, so our search is neither data-driven, nor would a successful theory of quantum gravity necessarily much expand the range of natural phenomena we can explain (or at least describe) conceptually and mathematically. A while ago, Chris Isham asked me rhetorically: ``What if someone came today, with a printout of three long articles in his or her hands, and claimed that these articles contain the Theory of Quantum Gravity? How would we judge if this person is right or wrong? Which criteria apply?''. 

In this article, I will consider some very general conceptual questions on the way to quantum gravity. Whilst these questions may seem metaphysical (a word that is often used in a pejorative sense by working physicists), in the end each technical approach will be confronted by such questions. I will make some remarks on the following questions:
\begin{enumerate}
	\item Is quantum theory necessarily quantum? Is it adequate to (try to) expand concepts of quantum ideas to a theory of quantum gravity, with the usual mathematical apparatus of quantum theory intact? Could quantum gravity be an instrumentalist (or operational) theory, or does it have to be a realist theory?
	\item If we aim at a realist form of quantum theory and theories `beyond quantum theory' such as quantum gravity, what kind of logic could we potentially use in the face of no-go theorems such as the Kochen-Specker theorem?
	\item In an encompassing theory of the whole universe, which r\^ole does the physicist play -- is she or he necessarily part of the description?
\end{enumerate}
Of course, I cannot hope to give full answers to these questions; I can merely sketch some ideas and recent technical developments using topos theory in physics which may become useful in finding answers to such questions. 

In section \ref{Sec_IsQGQuantum}, the well-known argument why instrumentalist interpretations are problematic in quantum gravity is presented. Yet, the conclusion we draw from this is non-standard: a theory of quantum gravity should not be based on Hilbert spaces. Section \ref{Sec_WhatLogic} gives an outline of some basic structures in the topos approach to the formulation of physical theories, leading to a neo-realist interpretation of the new topos-based mathematical formalism for quantum theory. The topos approach to the formulation of physical theories was initiated by Isham \cite{Ish97} and Isham/Butterfield \cite{IshBut98,IshBut99,IHB00,IshBut02}, was developed and substantially expanded by this author and Isham \cite{DoeIsh08a,DoeIsh08b,DoeIsh08c,DoeIsh08d,Doe09a,Doe09b,Doe11a,Doe11b,DoeIsh11,DoeIsh12,Doe12,Doe12b,Doe12c}, and further developed by Heunen, Landsman, Spitters and Wolters \cite{HLS09a,HLS09b,Wol10,HLS11,Woo11}, Flori and collaborators \cite{Flo11,BGeFlo12,BreFlo12} and others \cite{Nak11,Nui11}. In section \ref{Sec_Metalogic}, we briefly argue about the relation between `system logic' and metalogic, and section \ref{Sec_Concl} concludes.

\section{Is quantum gravity necessarily quantum?}			\label{Sec_IsQGQuantum}
This question may seem trivial at first sight: since quantum gravity is supposed to unify or reconcile quantum theory and general relativity, it will of course be some sort of quantum theory (just as it will also be some sort of theory of gravity). Yet, what is less clear is if a theory of quantum gravity necessarily must be based on the Hilbert space formalism?

Ever since von Neumann gave quantum mechanics its mathematical form in 1928, the Hilbert space formalism has been the mathematical underpinning of quantum theory. Further developments like quantum field theories added more mathematical structures, but the Hilbert space formalism remained the core of the mathematical apparatus of quantum theory.

\subsection{Interpretations of quantum theory}
Quantum theory, like every physical theory, consists of a mathematical apparatus and an interpretation that links the mathematics to physical entities and processes. In the case of quantum theory, a plethora of interpretations exists, and there is an ongoing debate about which of these interpretations is to be preferred. Importantly, the debate is largely concerned with interpretations of the Hilbert space formalism, while the Hilbert space formalism itself is rarely questioned. Hence, the underlying mathematical apparatus of quantum theory remains more-or-less fixed in this debate.

\textbf{Instrumentalism.} Most interpretations of quantum theory, in particular the classical Copenhagen interpretation and operational interpretations, which have become popular again recently, posit a fundamental distinction between quantum system and observer. Measurements are primitive notions and hence are not in need of a definition in such an interpretation. Necessarily, observers and their measuring devices are not quantum systems, but classical, which leads to many interpretational issues.

An instrumentalist interpretation does not give rise to a picture of reality. It restricts itself to predicting outcomes of experiments (often in a probabilistic sense) that an observer performs on the system. As such, an instrumentalist interpretation does not tell us much about what the quantum system `does' or `is' if we don't measure. Given a closed system, an instrumentalist view is not informative.

\textbf{Realism.} In contrast to an instrumentalist interpretation, a realist interpretation does not fundamentally depend on observers and measurements. Rather, such an interpretation gives a picture of reality, of `how things are' and what `is actually going on'. In a realist interpretation, observers and measurements are secondary concepts. Measurements can reveal what is going on, but they are not fundamentally adding anything. Of course, we idealise here and assume the case of non-disturbing measurements.

Yet, as is well known, it is very hard to come up with a realist interpretation of the Hilbert space formalism. The only two established examples are the de Broglie-Bohm pilot wave formalism, which has massive problems with the extension to special relativistic space-time, and the Everett many-worlds interpretation. The latter posits that whenever a quantum experiment with several possible outcomes is performed, all outcomes are realised and the universe splits up into corresponding branches. This (not very frugal) ontology may be acceptable for some philosophers of physics, but it does not seem attractive to us. What is `real' in many-worlds is the wave function of the universe, which of course is inaccessible in principle and does not undergo measurement or collapse. It is doubtful if this can be seen as a `picture of reality' of the kind we are aiming at. An observer in many-worlds only has experience of one branch, but we also have to take the god's-eye view of the wave function of the universe to make sense of the theory.

\textbf{Born rule and instrumentalism.} In fact, the Hilbert space formalism almost forces an instrumentalist interpretation on us. A key aspect of the Hilbert space formalism, and the link to observable phenomena, is the Born rule that allows calculating expectation values of observables when the system is in a given state. Of course, the concepts of expectation values and probabilities presuppose the two-level ontology system-observer and are dependent on (repeated) measurements. Hence, there is a direct link between the usual interpretation of the Born rule and instrumentalist interpretations of quantum theory. It is much-debated whether a many-worlds interpretation, in which every possible outcome occurs and is equally `real', can reproduce the probabilistic predictions of quantum theory.

\textbf{Relativistic quantum field theories.} It is commonly accepted that the interpretational problems of non-relativistic quantum theory are not solved by going to relativistic quantum field theory. Instead of expectation values, one considers cross sections, scattering probabilities, etc., but also these arise from experiments performed by classical observers. This means that at least implicitly we use an instrumentalist interpretation also in QFT.

\subsection{Instrumentalism in quantum gravity?}
It is obvious that a formulation of quantum gravity based on Hilbert spaces would inherit the interpretational problems associated with instrumentalism, since a mathematical apparatus based on Hilbert spaces most naturally combines with an instrumentalist interpretation to give a physical theory. Yet, there are strong reasons to doubt the usefulness of instrumentalist interpretations in quantum gravity.

\textbf{Problems with instrumentalism in quantum gravity.} Firstly, if we assume that quantum gravity, like its classical counterpart, is a theory of the whole universe, then there is no external observer who could perform measurements on this system. As mentioned above, instrumentalist interpretations are not very useful for closed systems. The universe is the ultimate (and only true) closed system.

A second reason to doubt the usefulness of instrumentalist interpretations in quantum gravity is that the concept of measurements seems to presuppose a space-time background, since measurements take place at some location at some point (or during some period) in time. It has often been argued that quantum gravity should be formulated in a background-independent way, but measurement does not seem to be background-free notion. In a theory of quantum gravity, presumably space and time will be treated as quantum objects, whatever that will mean in detail. What could a measurement of quantum space or quantum time mean -- \emph{where} and \emph{when} would such a measurement take place?

\textbf{Quantum gravity without Hilbert spaces.} These issues seem serious enough to doubt that any instrumentalist (or operational) interpretation could be useful in quantum gravity. If, moreover, we take into account the fact that any theory based on the mathematical apparatus of Hilbert spaces practically forces an instrumentalist interpretation on us, we come to the following conclusion:
\begin{center}
\emph{The mathematical apparatus of a future theory of quantum gravity should not be based on Hilbert spaces and hence should not be a quantum theory in the standard sense. }
\end{center}
Instead, we should try to formulate a theory of quantum gravity in such a way that the mathematical apparatus can be combined with a realist interpretation, avoiding the serious conceptual issues with instrumentalist interpretations sketched above. This means we should strive for a mathematically and conceptually new form of theory of quantum gravity, departing from the paradigm of Hilbert spaces.

Naturally, a good starting point for such a project is not quantum gravity, but much more modest non-relativistic quantum theory. What kind of mathematical apparatus, replacing the Hilbert space formalism, is there that would allow a realist interpretation in a natural manner? We emphasise that this is \emph{not} asking for yet another interpretation of the established formalism, but much more radically for a mathematical re-formulation of quantum theory, together with a conceptually new, realist kind of interpretation of this new mathematical formalism.

\textbf{The topos approach.} The topos approach to the formulation of physical theories, and in particular to quantum theory \cite{DoeIsh11}, is an attempt to provide such a mathematical reformulation of quantum theory, together with a new, realist interpretation. The technical details are involved and can be found elsewhere. Here, we focus on some conceptual ingredients and particularly focus on some logical aspects.

\section{What logic for a realist form of quantum theory?}			\label{Sec_WhatLogic}
\subsection{Realism in classical physics}
If we aim to be realists, the first question is: realists about what? The prototype of a realist theory is classical mechanics. At a very basic level, this is a theory based on a state space, a space of values of physical quantities (which is the real numbers), and physical quantities as maps from the state space to the space of values. A physical quantity $A$, for example position, is represented by a real-valued function $f_A$ from the state space, given mathematically by some set $\cS$ (typically a symplectic or Poisson manifold), to the space of values, given mathematically by the real numbers $\bbR$.

If we consider a subset $\De$ of the real line, then $f_A^{-1}(\De)$ is a subset of the state space $\cS$. This subset represents a \emph{proposition} ``$\Ain\De$'', that is, ``the physical quantity $A$ has a value in the set $\De\subseteq\bbR$''. The subset $f_A^{-1}(\De)$ consists of all those states, i.e., elements of the state space $\cS$, for which the proposition is $true$. If the state $s$ of the classical system is contained in $f_A^{-1}(\cS)$, then the physical quantity $A$ has a value in the set $\De$. Otherwise, if $s\notin f_A^{-1}(\De)$, then $A$ does not have a value in $\De$, and the proposition ``$\Ain\De$'' is $false$.

Different propositions such as ``$\Ain\De$'', ``$B\varepsilon\Ga$'', ``$C\varepsilon\Xi$'' about the values of physical quantities correspond to (generally) different subsets $f_A^{-1}(\De)$, $f_B^{-1}(\Ga)$, $f_C^{-1}(\Xi)$ of the state space $\cS$. The \emph{conjunction} ``$\Ain\De$ and $B\varepsilon\Ga$'' corresponds to the intersection $f_A^{-1}(\De)\cap f_B^{-1}(\Ga)$, the \emph{disjunction} ``$\Ain\De$ or $B\varepsilon\Ga$'' corresponds to the union $f_A^{-1}(\De)\cup f_B^{-1}(\Ga)$, and the \emph{negation} ``not $\Ain\De$'' corresponds to the complement $\cS\backslash f_A^{-1}(\De)$.

Hence, in classical physics there is an algebra of propositions, with conjunction, disjunction and negation, and this algebra is represented mathematically by the Boolean algebra $\mc P(\cS)$ of subsets of the state space $\cS$. Moreover, states act as \emph{models} for this propositional theory, i.e., they assign truth values to propositions. Mathematically, each point $s$ of the state space gives a map
\begin{align}
				t_s:\mc P(\cS) &\lra (false,true)\\
				X &\lmt \left\{\begin{tabular}[c]{ll}%
												$true$ & if $s\in X$\\
												$false$ & if $s\notin X$.
											\end{tabular}
								\ \right.
\end{align}
Clearly, $(false,true)$ is a Boolean algebra itself, and $t_s$ is a morphism of Boolean algebras, that is, $t_s(X\cap Y)=t_s(X)\wedge t_s(Y)$ etc. 

Classical physics is a realist theory in the sense that 
\begin{itemize}
	\item [(a)] There is a space of states $\cS$ whose subsets are interpreted as representatives of propositions of the form ``$\Ain\De$''.
	\item [(b)] The subsets of $\cS$ form a Boolean algebra $\mc P(\cS)$. The algebraic operations $\cap,\cup$ and $\cS\backslash\_$ represent the logical operations of conjunctions, disjunctions and negations of propositions.
	\item [(c)] States $s\in\cS$ provide models of the propositional theory represented by $\mc P(\cS)$, i.e., they are Boolean algebra morphisms from $\mc P(\cS)$ to the Boolean algebra $(false,true)$ of truth values.
	\item [(d)] Every proposition $X\subseteq\mc P(\cS)$ has a truth value $t_s(X)$ in every given state $s\in\cS$, and every physical quantity $A$ has a value $f_A(s)\in\bbR$ in a given state $s$.
\end{itemize}
Classical physics is realist about propositions and their truth values. There is a `way things are', independent from observers and measurements. We want to take this as the model for more general realist theories.

Yet, the Kochen-Specker theorem \cite{KocSpe67,Doe05} seems to pose a strict limitation on any attempt at providing a realist form of quantum theory in this sense: it shows that under weak and natural assumptions, there is no way of assigning truth values to all propositions like ``$\Ain\De$'' in a consistent way. Mathematically, there is no way of embedding the algebra representing propositions about a quantum system into a Boolean algebra. 

\subsection{The topos approach: from sets to presheaves, from Boolean logic to intuitionistic logic}
The Kochen-Specker no-go theorem can be circumvented by relaxing the assumptions we make. In particular, we can (a) allow the representatives of propositions to form a weaker structure than a Boolean algebra, (b) allow more truth values than just $true$ and $false$, and (c) allow more general maps than Boolean algebra morphisms as states or models of our propositional theory. 

If, after relaxing assumptions in this way, we have a theory in which all propositions have truth values in all given states, then we still regard this as a mathematical formalism that can be interpreted in a realist way. Of course, we have to show that quantum theory can be re-formulated in such a way.

\textbf{Some ingredients of the topos approach to quantum theory.} As mentioned above, the topos approach to quantum theory provides such a mathematical reformulation of quantum theory, together with a realist interpretation. We briefly sketch the main ingredients of the mathematical apparatus.

The topos approach gives a state space picture of quantum theory in strong analogy to the state space picture of classical physics. First of all, there is a notion of state space. Yet, this object is not assumed to be a set, but is a more general kind of object. Concretely, we use a \emph{presheaf}, i.e., a varying set, as will be explained in more detail below. Also the space of values is a presheaf (of real intervals), not just the set of real numbers as in classical physics. In analogy to classical physics, physical quantities are represented by maps from the state presheaf to the value presheaf. These maps are not mere functions, but maps between presheaves (natural transformations). Moreover, propositions such as ``$\Ain\De$'' are represented by sub`sets' (in fact, subpresheaves) of the state presheaf. Finally, states are not represented by points of the state presheaf -- it turns out the state presheaf has no points at all in a suitable technical sense! This is exactly equivalent to the Kochen-Specker theorem. Instead of points, one uses certain minimal (i.e., small) subobjects of the spectral presheaf to represent pure states.

\textbf{Contexts and partial world views.} A fundamental feature of quantum theory is that experimentally, we only have partial access to the system in the sense that only certain, compatible physical quantities can be measured simultaneously. Mathematically, these are represented by commuting self-adjoint operators, forming a commutative subalgebra of the algebra of physical quantities. Such a subalgebra, and the partial perspective on the quantum system that it describes, is called a \emph{context}.

It was Bohr's doctrinal view that one should only speak about quantum systems in classical terms. This basically amounts to picking out a single context by a measurement setup, after which it becomes impossible to speak meaningfully about the values of physical quantities not contained in this context. We follow a radically different route here: instead of singling out a particular context, we consider \emph{all} of them simultaneously and treat them on equal footing. We collect all the partial perspectives on a quantum system. Mathematically, we consider the set of all commutative subalgebras of a noncommutative algebra $\cN$ of physical quantities, and we partially order this set by inclusion. This poset (partially ordered set) is called the context category and is denoted $\VN$.

While it may look very simple-minded to cut a noncommutative algebra into commutative pieces, the context category contains a surprising amount of information about the original noncommutative algebra: for the case of a von Neumann algebra $\cN$, one can show that the context category $\VN$ determines the original algebra up to Jordan isomorphisms \cite{HarDoe10}. The formalism becomes powerful because we keep track of how contexts overlap, i.e., intersect.

\textbf{The spectral presheaf, subobjects and propositions.} The state object for quantum theory is the so-called \emph{spectral presheaf $\Sig$}. This is a varying set over the context category $\VN$. To each context $V\in\VN$, we assign the Gelfand spectrum $\Sig_V$ of the algebra, which is a compact Hausdorff space. This space can be seen as a `local state space' for the physical quantities contained in the context $V$, where `local' means `at this context within the global noncommutative algebra'. If $V'\subset V$ is a smaller context, then there is a canonical continuous, surjective function from $\Sig_V$, the Gelfand spectrum of the bigger algebra, to $\Sig_{V'}$, the spectrum of the smaller algebra. In this way, we `glue together' all the local state spaces into a global object $\Sig$, the spectral presheaf, which is the state object for quantum theory. One can show that the spectral presheaf has no global elements \cite{IshBut98,IHB00,Doe05}, which are the presheaf analogues of points. This lack of points is equivalent to the Kochen-Specker theorem.

Being a presheaf, $\Sig$ is an object in the \emph{topos $\SetVNop$ of presheaves over $\VN$}. By a standard result, the subobjects, that is, subpresheaves, of any object in a topos form a \emph{Heyting algebra}. Just as Boolean algebras mathematically represent classical Boolean propositional logics, Heyting algebras represent intuitionistic logics, which are more general than Boolean logics, because the law of the excluded middle need not hold. 

In particular, the subobjects of the spectral presheaf $\Sig$ form a Heyting algebra, and we use this structure to represent propositions about the values of physical quantities of our quantum system. There is a map from propositions of the form ``$\Ain\De$'' to subobjects of the spectral presheaf, called \emph{daseinisation of projections} that was discussed in detail in \cite{DoeIsh08b,Doe11b}.

\textbf{The value presheaf and representation of physical quantities.} In the topos approach, physical quantities do not take values in the usual real numbers (which would run into trouble with the Kochen-Specker theorem). Instead, we allow more general, `unsharp' values in the form of real intervals. The value presheaf is denoted $\Rlr$.

Physical quantities are represented by arrows (natural transformations) from $\Sig$ to $\Rlr$. The eigenstate-eigenvalue link is preserved in the sense that if one has an eigenstate of some physical quantity $A$, the value assigned to this physical quantity at all contexts that contain $A$ is the one-point interval $[a,a]$ that just contains the eigenvalue $a$. Details can be found in \cite{DoeIsh08c,Doe11b,DoeBar12}. In order to represent propositions such as ``$\Ain\De$'', instead of using daseinisation of projections, one can also consider inverse images (technically, pullbacks along arrows in the topos) of subobjects of the presheaf $\Rlr$ to obtain subobjects of the spectral presheaf $\Sig$, see sections 13.8.7--8 in \cite{DoeIsh11}. This is structurally analogous to the situation in classical physics, where a proposition ``$\Ain\De$'' is represented by the subset $f_A^{-1}(\De)$ of the state space.

\textbf{Topos logic and neo-realism.} Crucially, every topos comes with a built-in logic. This is a higher-order, typed, intuitionistic logic, often multi-valued. For the general theory, see \cite{McLMoe92,Joh02/03}.

In our case, the topos associated with a quantum system is $\SetVNop$, presheaves over the context category. The available truth values in the logic of this topos are lower sets in the context category $\VN$: subsets $T\subseteq\VN$ such that if $V\in T$ and $V'\subset V$, then $V'\in T$. There are uncountably many truth values available instead of just $true$ and $false$, and the truth values form a Heyting algebra themselves.

Recall that in classical physics, a proposition ``$\Ain\De$'' is represented by a subset $f_A^{-1}(\De)$ of the state space $\cS$, and a state is represented by a point $s\in\cS$. The truth value of the proposition in the state is the truth value of a the Boolean formula $s\in f_A^{-1}(\De)$, that is,
\begin{align}
			v(\text{``}\Ain\De\text{''};s)=(s\in f_A^{-1}(\De))\in (false,true).
\end{align}
In the topos approach, a proposition ``$\Ain\De$'' is represented by a subobject $\ps\deo(\hP)$ of the state object, the spectral presheaf $\Sig$. A pure state $|\psi\rangle$ is represented not by a point of $\Sig$ (there are none), but by a certain minimal subobject $\wpsi$. We can interpret the formula $\wpsi\in\ps\deo(\hP)$ in the internal logic of our topos using the so-called Mitchell-Benabou language, which gives a truth value,
\begin{align}
			v(\text{``}\Ain\De\text{''};\wpsi)=(\wpsi\in\ps\deo(\hP)).
\end{align}
In the classical case, a point $s$ either lies in a subset $f_A^{-1}(\De)$ (giving $true$ globally) or not (giving $false$ globally). In the topos, we do not just get a single $true$ or $false$, but one such truth value for each context $V\in\VN$. The actual topos truth value is the collection of all these `local' truth values. It is easy to show that if we have $true$ at $V$ and $V'\subset V$, then we also get $true$ at $V'$. Globally, we get a lower set in $\VN$ and hence a truth value in the intuitionistic, contextual and multi-valued logic of our topos.

Just as in classical physics, every proposition has a truth value in any given state. There is no fundamental reference to observers, measurements or other instrumentalist concepts. The topos approach provides a mathematical formalism for quantum theory that can naturally be given a realist interpretation. The price to pay is that the logic employed is not classical two-valued Boolean logic, but the intuitionistic, multi-valued logic provided by the topos of presheaves. For this reason, we usually speak of a \emph{neo-realist} interpretation.

Many other aspects of quantum theory can be described in the new topos-based mathematical formalism, such as mixed states \cite{Doe09a,Doe11a}, time evolution \cite{Doe12c}, probabilities and the Born rule \cite{Doe09a,DoeIsh12,FRV12}, etc.

\textbf{Extending to field theories and beyond.} Whilst we have considered only non-relativistic systems described by an noncommutative algebra of physical quantities, the general scheme can be extended straightforwardly to other kinds of theories. For example, instead of just taking the poset $\VN$ of contexts, one can also consider systems where subalgebras of observables are attached to space-time regions, as in algebraic quantum field theory. A suitable context category would then also carry additional space-time labels. First steps in this direction were taken by Joost Nuiten in \cite{Nui11}.

Even more generally, the basic scheme of a state space, a space of values and physical quantities as maps between them is so general that it applies virtually to all physical theories. By modelling state spaces and spaces of values as objects in a topos, e.g. as presheaves or sheaves, and physical quantities as arrows in the topos between them, many generalisations beyond the usual set-based and Hilbert space-based mathematics become available. In each case, the topos comes with a built-in intuitionistic logic, and logical formulas (e.g. about the values of physical quantities in a given state) can be interpreted within the logic of the topos, in a manner completely analogous to the one sketched above for quantum theory. 

Crucially, all propositions have truth values in any given state, without the need to invoke observers or measurements -- there is a natural neo-realist interpretation of the topos formalism, no matter what the specific topos is and what the choices of the state object and value object are. In this way, the topos approach to the formulation of physical theories avoids the serious interpretational issues that instrumentalist interpretations of the Hilbert space formalism have.

\section{A logic for physical systems and a metalogic for physics}			\label{Sec_Metalogic}
In this short section, I present some quite speculative thoughts on the r\^oles of the topos-internal logic and the topos-external metalogic in which we define the topos and structures within it.

The internal logic of a topos generally is intuitionistic (only in particular cases, it is Boolean), which means that the law of the excluded middle does not hold. A topos can be seen as a generalised universe of sets and hence as an arena to do mathematics. Proofs in a topos with an intuitionistic internal logic must necessarily be constructive, since proof by contradiction is not available. Moreover, typically the axiom of choice is not available in a given topos, though weaker forms like countable choice may hold. 

If we use a topos and its internal logic to argue about physical systems, it seems that we commit ourselves to using constructive mathematics. There is interesting work along these lines by Heunen, Landsman, Spitters \cite{HLS09a,HLS09b,Wol10,HLS11} and Fauser, Raynaud and Vickers \cite{FRV12}.

Yet, when we do physics, we necessarily have to separate ourselves from the system to be described. We do not aim at providing some sort of inclusive report, but rather try to give an objective or at least inter-subjective description of some system or phenomenon outside of us. There may well be a logic adequate to the system in itself, but this system logic is not the logic in which we are thinking and arguing \emph{about} the system. For this, we use a metalogic, typically Boolean, in which we define the mathematical structures to describe the system. For this reason, in our description and mathematical arguments about the system we are free to use the metalogic. Of course, we must take care not to mix metalogical arguments about the system with arguments within the logic of the system.

In the topos approach, we define a topos and certain mathematical structures in it, e.g. the state object. This definition takes place in a (typically Boolean) metalogic about which we do not reflect very much. As we showed, the internal logic of the topos is useful in quantum theory, in the sense that propositions about the values of physical quantities can be assigned truth values using this `system logic'. We cannot use the Boolean metalogic to do this assignment of truth values, but we have to reflect about the physical interpretation of such truth values from the external, metalogical perspective. 

In short, our metalogic for doing physics is not the same as the system logic provided by the topos in which we describe a given physical system mathematically.

This even applies if the physical system we consider is the whole universe, as in a theory of quantum gravity: if we argue physically, then we are outside of the system we describe. Of course, even if we argue about the whole universe, we actually only consider a very small number of degrees of freedom. There is no (meta)logical contradiction arising, we can describe the whole universe but still `step out' of it when doing so, since our description is very far from complete.

\section{Conclusion}			\label{Sec_Concl}
In this contribution, we gave some conceptual arguments concerning the `logical shape' of a future theory of quantum gravity. We first presented the well-known argument that quantum gravity will not be a theory whose mathematical apparatus comes with an instrumentalist interpretation, since such an interpretation makes no sense for closed systems with no external observers. Moreover, we argued that any theory based on Hilbert spaces almost automatically comes with an instrumentalist interpretation, which led us to the radical conclusion that a theory of quantum gravity should not be based on Hilbert spaces. 

Instead, we suggest to use a form of theory that is based on a state space picture, generalising from classical physics. Such theories more naturally lend themselves to realist interpretations. We sketched some aspects of the topos approach to quantum theory and argued that by going from sets to presheaves and from Boolean logic to intuitionistic logic, we arrive at a mathematical formalism for quantum theory that has a natural neo-realist interpretation. Moreover, the underlying scheme is general enough to allow generalisations to field theories and beyond. Needless to say, much work remains to be done.

Finally, we briefly argued that even when we commit ourselves to describing the whole universe using structures in a topos, and if we use the internal logic of the topos to assign truth values to propositions etc., we do not have to do all our proofs and mathematical arguments internally in the topos, i.e., constructively. Doing physics necessarily means to separate oneself from the system to be described, even if this system is the whole universe. Since we have to `step out' of the system, we have to argue using the (typically Boolean) metalogic in which we define the mathematical structures, e.g. topoi and state objects, that we use in the mathematical description of the system at hand. It is this Boolean metalogic in which we do physics.

\textbf{Acknowledgements.} Chris Isham has been the major influence on my physical thinking over the last few years, and I very much thank him for numerous discussions and his friendship. Dicussions with Harvey Brown, Steve Vickers, Klaas Landsman, Masanao Ozawa and Tim Palmer are gratefully acknowledged. I thank my students Carmen Constantin, Rui Soares Barbosa, Dan Marsden and Nadish de Silva for their questions and their valuable input. Finally, I thank Dean Rickles for inviting me to contribute to this volume.


\begin{thebibliography}{1}

\bibitem{Aud79} W.H.~Auden, \textit{Selected Poems}, edited by E. Mendelson, faber and faber (1979).

\bibitem{vdBHeu10} B.~van den Berg, C.~Heunen, ``Noncommutativity as a colimit'', arXiv:1003.3618v3 (2010; version 3 from 25. April 2012).

\bibitem{BGeFlo12} J.~Ben Geloun, C.~Flori, ``Topos Analogues of the KMS State'', arXiv:1207.0227v2 (2012).

\bibitem{BreFlo12} W.~Brenna, C.~Flori, ``Complex Numbers, One-Parameter of Unitary Transformations and Stone's Theorem in Topos Quantum Theory '', arXiv:1206.0809v2 (2012).

\bibitem{Doe05} A.~D\"{o}ring, ``Kochen-Specker theorem for von Neumann algebras'', \textit{Int. Jour. Theor. Phys.} {\bf 44}, 139--160 (2005).

\bibitem{Doe09a} A.~D\"{oring}, ``Quantum States and Measures on the Spectral Presheaf'', \emph{Adv. Sci. Lett.} \textbf{2}, special issue on ``Quantum Gravity, Cosmology and Black Holes'', ed. M. Bojowald, 291--301 (2009).

\bibitem{Doe09b} A.~D\"{o}ring, ``Topos theory and `neo-realist' quantum theory'', in \textit{Quantum Field Theory, Competitive Models}, eds. B. Fauser, J. Tolksdorf, E. Zeidler, Birkh\"auser, Basel, Boston, Berlin (2009).

\bibitem{Doe11a} A.~D\"{o}ring, ``Topos quantum logic and mixed states'', in \textit{Proceedings of the 6th International Workshop on Quantum Physics and Logic (QPL 2009)}, \textit{Electronic Notes in Theoretical Computer Science} \textbf{270}, No. 2 (2011).

\bibitem{Doe11b} A.~D\"oring, ``The Physical Interpretation of Daseinisation'', in \textit{Deep Beauty}, ed. Hans Halvorson, Cambridge University Press, Cambridge, 207--238 (2011).

\bibitem{Doe12} A.~D\"oring, ``Topos-based Logic for Quantum Systems and Bi-Heyting Algebras'', to appear in \textit{Logic \& Algebra in Quantum Computing}, Lecture Notes in Logic, Association for Symbolic Logic in conjunction with Cambridge University Press; arXiv:1202.2750 (2012).

\bibitem{Doe12b} A.~D\"oring, ``Generalised Gelfand Spectra of Nonabelian Unital $C^*$-Algebras I: Categorical Aspects, Automorphisms and Jordan Structure'', arXiv (2012).

\bibitem{Doe12c} A.~D\"oring, ``Generalised Gelfand Spectra of Nonabelian Unital $C^*$-Algebras II: Flows and Time Evolution of Quantum Systems'', arXiv (2012).

\bibitem{DoeBar12} A.~D\"oring, R. Soares Barbosa, ``Unsharp Values, Domains and Topoi'', in \textit{Quantum Field Theory and Gravity, Conceptual and Mathematical Advances in the Search for a Unified Framework}, eds. F. Finster et al., Birkh\"auser, Basel, 65--96 (2012).

\bibitem{DoeDew12a} A.~D\"oring, B. Dewitt, ``Self-adjoint Operators as Functions I: Lattices, Galois Connections, and the Spectral Order'', arXiv:1208.4724 (2012).

\bibitem{DoeDew12b} A.~D\"oring, B. Dewitt, ``Self-adjoint Operators as Functions II: Quantum Probability'', arXiv:1210.5747 (2012).

\bibitem {DoeIsh08a} A.~D\"{o}ring, C.J. Isham, ``A topos foundation for theories of physics: I. Formal languages for physics'', \emph{J. Math. Phys.} \textbf{49}, Issue 5, 053515 (2008).

\bibitem {DoeIsh08b} A.~D\"{o}ring, C.J. Isham, ``A topos foundation for theories of physics: II. Daseinisation and the liberation of quantum theory'', \emph{J. Math. Phys.} \textbf{49}, Issue 5, 053516 (2008).

\bibitem {DoeIsh08c} A.~D\"{o}ring, C.J. Isham, ``A topos foundation for theories of physics: III. Quantum theory and the representation of physical quantities with arrows \mbox{$\breve\delta(\hat A):\Sig\ra\ps{\bbR^{\leftrightarrow}}$}'', \emph{J. Math. Phys.} \textbf{49}, Issue 5, 053517 (2008).

\bibitem{DoeIsh08d} A.~D\"{o}ring, C.J. Isham, ``A topos foundation for theories of physics: IV. Categories of systems'', \emph{J. Math. Phys.} \textbf{49}, Issue 5, 053518 (2008).

\bibitem{DoeIsh11}  A.~D\"{o}ring, and C.J. Isham, ```What is a Thing?': Topos Theory in the Foundations of Physics'', in \emph{New Structures for Physics}, ed. B.~Coecke, Lecture Notes in Physics \textbf{813}, Springer, Heidelberg, Dordrecht, London, New York, 753--937 (2011).

\bibitem {DoeIsh12} A.~D\"oring, C.J.~Isham, ``Classical and Quantum Probabilities as Truth Values'', \textit{J. Math. Phys.} \textbf{53}, 032101 (2012).

\bibitem{FRV12} B.~Fauser, G.~Raynaud, S.~Vickers, ``The Born rule as structure of spectral bundles (extended abstract)'', arXiv:1210.0615 (2012).

\bibitem{Flo11} C.~Flori, ``Group Action in Topos Quantum Physics'', arXiv:1110.1650v3 (2011).

\bibitem{HarDoe10} J.~Harding, A.~D\"oring, ``Abelian subalgebras and the Jordan structure of a von Neumann algebra'', arXiv:1009.4945 (2010).

\bibitem{HLS09a} C.~Heunen, N.P.~Landsman, B.~Spitters, ``A topos for algebraic quantum theory'', \emph{Comm.\ Math.\ Phys.} \textbf{291}, 63--110 (2009).

\bibitem{HLS09b} C.~Heunen, N.P.~Landsman, B.~Spitters, ``Bohrification of von {N}eumann algebras and quantum logic'', \textit{Synthese}, online first, DOI: 10.1007/s11229-011-9918-4 (2011).

\bibitem{HLS11} C.~Heunen, N.P.~Landsman, B.~Spitters, ``Bohrification'', in \textit{Deep Beauty}, ed. H.~Halvorson, Cambridge University Press, 271--313 (2011).

\bibitem{Ish97} C.J. Isham, ``Topos theory and consistent histories: The internal logic of the set of all consistent sets'', {\em Int. J. Theor. Phys.}, \textbf{36}, 785--814 (1997).

\bibitem{IshBut98} C.J.~Isham, J.~Butterfield, ``A topos perspective on the Kochen-Specker theorem: I. Quantum states as generalised valuations'', \textit{Int. J. Theor. Phys.} \textbf{37}, 2669--2733 (1998).

\bibitem{IshBut99} C.J.~Isham and J.~Butterfield, ``A topos perspective on the {K}ochen-{S}pecker theorem: II. Conceptual aspects, and classical analogues'', \textit{Int.\ J.\ Theor.\ Phys.} \textbf{38}, 827--859 (1999).

\bibitem{IHB00} C.J.~Isham, J.~Hamilton, J.~Butterfield, ``A topos perspective on the Kochen-Specker theorem: III. Von Neumann algebras as the base category'', \textit{Int. J. Theor. Phys.} \textbf{39}, 1413--1436 (2000).

\bibitem{IshBut02} C.J.~Isham, J.~Butterfield, ``A topos perspective on the Kochen-Specker theorem: IV. Interval valuations'', \textit{Int.\ J.\ Theor.\ Phys} \textbf{41}, 613--639 (2002).

\bibitem{Joh02/03} P.T.~Johnstone, \textit{Sketches of an Elephant: A Topos Theory Compendium}, Vols. 1\&2, Oxford Logic Guides \textbf{43}\&\textbf{44}, Oxford University Press, Oxford (2002/03).

\bibitem {KocSpe67} S.~Kochen, E.P.~Specker, ``The problem of hidden variables in quantum mechanics'', \textit{Journal of Mathematics and Mechanics} \textbf{17}, 59--87 (1967).

\bibitem{McLMoe92} S.~Mac{L}ane, I.~Moerdijk, \textit{Sheaves in Geometry and Logic: A First Introduction to Topos Theory}, Springer, New York, Berlin, Heidelberg (1992).

\bibitem{Nak11} K.~Nakayama, ``Sheaves in Quantum Topos Induced by Quantization'', arXiv:1109.1192v2 (2011).

\bibitem{Nui11} J.~Nuiten, ``Bohrification of local nets of observables'', Bachelor thesis, Radboud University Nijmegen, arXiv:1109:1397 (2011).

\bibitem{Wit63} L.~Wittgenstein, \textit{Tractatus logico-philosophicus: Logisch-philosophische Abhandlung}, Suhrkamp (1963)

\bibitem{Wol10} S.~Wolters, ``A Comparison of Two Topos-Theoretic Approaches to Quantum Theory'', arXiv:1010.2031v2 (version 2 from 3. August 2011).

\bibitem{Woo11} T.~Woodhouse, ``Time Evolution in Quantum Theory and Quantum Information, A Topos Theoretic Perspective'', MSc thesis, University of Oxford (2011).

\end{thebibliography}
\end{document}